\documentstyle[12pt,aaspp4,flushrt]{article}

\def\Msol{\mbox{M$_\odot$}}
\def\Lsol{\mbox{L$_\odot$}}
\def\etal{{\it et~al.~}}
\def\eg{{\it e.g.~}}

\def\degrees{\mbox{$^{\circ}$}}
\def\ha{H$\alpha$/[NII]~}
\def\Sec{\rlap{$^{\prime\prime}$}.\hbox to 2pt{}}
\def\Min{\rlap{$^\prime$}.\hbox to 1pt{}}

\begin{document}

\title{The Nuclear Region of M51 Imaged with the
{\it HST} Planetary Camera\footnote[1]{Based 
on observations with the NASA/ESA Hubble Space Telescope, obtained at the
Space Telescope Science Institute, which is operated by AURA, Inc., under
NASA contract NAS 5-26555.}}

\author{Carl J. Grillmair}
\affil{Jet Propulsion Laboratory, California Institute of Technology,
4800 Oak Grove Drive, Pasadena, California 91109-8099}

\author{S. M. Faber}
\affil{UCO/Lick Observatory, Board of Studies in Astronomy and
Astrophysics, University of California, Santa Cruz, California
95064}

\author{Tod R. Lauer}
\affil{Kitt Peak National Observatory, National Optical Astronomy
Observatories\footnote[2]{The National Optical Astronomy Observatories are
operated by the Association of Universities for Research in Astronomy,
Inc. (AURA) under cooperative agreement with the National Science
Foundation.}, P.O. Box 26732, Tucson, Arizona 85726}

\author{J. Jeff Hester}
\affil{Department of Physics and Astronomy, Arizona State University,
Tempe, Arizona 85287-1504}
	
 \author{C. Roger Lynds} \affil{Kitt Peak National Observatory,
National Optical Astronomy Observatories\footnotemark[2], P.O. Box
26732, Tucson, Arizona 85726}

\author{Earl J. O'Neil, Jr.} \affil{Kitt Peak National Observatory,
National Optical Astronomy Observatories\footnotemark[2], P.O. Box
26732, Tucson, Arizona 85726}

\author{Paul A. Scowen}
\affil{Department of Physics and Astronomy, Arizona State University,
Tempe, Arizona 85287-1504}

\begin{abstract}

	 We present high-resolution, broad- and narrow-band,
pre-refurbishment images of the central region of M51 taken with the
Planetary Camera of the {\it Hubble Space Telescope}. The $V$-band
images show a rather chaotic distribution of dust lanes, though some
are oriented radially, roughly aligned with the major axis of
the bar, and may be transporting gas to the AGN in the nucleus. The
dust lane obscuring the nucleus of the galaxy, which was previously
thought to be an edge-on accretion disk feeding the AGN, is not
centered on the nucleus. It is unlikely that this is a stable
configuration, suggesting that the material has only recently entered
the nuclear region.  The nucleus is contained within a cluster of
stars having a total luminosity of order $5 \times 10^7
\Lsol$. Fitting a King model to the least obscured portions of the
cluster yields a maximum core radius of 14 pc.  The morphology
apparent in the forbidden-line images of the extra-nuclear cloud is
consistent with a narrow jet striking and scattering off
the boundary of a relatively dense cocoon of gas in the disk of the
galaxy. The emission-line regions are concentrated along the inner
borders of dust filaments, supporting the view that the nuclear jet
is ramming into and stirring up the ISM of the disk.

\end{abstract}


\section{Introduction}

	M51 (NGC 5194, the ``Whirlpool'' galaxy) is a relatively
nearby, nearly face-on, Sc spiral. Its proximity, combined with
grand-design spiral arms and low-level nuclear activity have made it a
popular target for both ground-based and space-based instrumentation.
Goad, De Veny, \& Goad \markcite{gdvg79} (1979) first determined that
M51 has a Seyfert-like nucleus, and Rose and Cecil \markcite{rc83}
(1983) measured line widths at zero intensity of $\sim 1800$ km
s$^{-1}$. Ford \etal \markcite{fcjlh85} (1985, hereafter FCJLH) showed
that much of this activity originates outside the nucleus,
specifically in a filled cloud to the south and a ring of
emission-line regions to the north of the nucleus.  Based on the high
temperature of the gas in the cloud and the similarity of the
emission-line ratios to those found in supernova remnants, they
determined that the emission-line gas is probably excited by shocks
rather than photoionization by a non-stellar source in the nucleus.
The morphology and kinematics of the southern cloud led FCJLH to
conclude that it was an expanding shell of gas in the disk of the
galaxy, either inflated by a relativistic jet emanating from the
nucleus, or ejected from the nucleus as a distinct plasmoid.  These
results were later confirmed and extended by Cecil \markcite{c88}
(1988), who used imaging Fabry-Perot observations to map the
narrow-line region and presented a detailed model of a nuclear jet
impinging on gas in the disk.  Spencer \& Burke \markcite{sb72} (1972)
discovered a weak, elongated nuclear radio source in M51, and Crane \&
van der Hulst \markcite{cvh92} (1992) used sub-arcsecond 6cm VLA
observations to show that the extra-nuclear cloud and the nucleus are
indeed connected by a narrow ($\le 0.3\arcsec$) jetlike radio feature.
However, the energy in the jet is less than that required to account
for the total number of recombinations, and a non-thermal, nuclear
source of ionizing radiation appears necessary.

	 {\it Hubble Space Telescope} (HST) images of the central
regions of M51 were first taken in 1991, and these were briefly
discussed by Kinney \etal \markcite{k92} (1992) and Mitton \markcite{m92}
(1992). Most striking was the F547M image, which showed an
``X''-shaped dust feature obscuring the nucleus. The main East-West
bar of the ``X'' was roughly perpendicular to the radio jet, leading
some to speculate that we were looking edge-on at a torus of material
surrounding and presumably feeding the massive black hole assumed to
be powering the jet.

	As part of an ongoing GTO project to study the physical
conditions in the nuclei of relatively nearby galaxies, we have
acquired high-resolution, $V$-band Planetary Camera images of the
central region of M51. We present and briefly discuss these images,
along with complementary narrow-band images from the HST archive.

\section{Observations. \label{sec:obs}}

	 Planetary Camera (PC) images of the nuclear region of M51
were taken in July of 1992 using filter F555W (roughly corresponding
to the Johnson $V$ band) as part of Wide Field/Planetary Camera (WF/PC
I) Guaranteed Time Observer program 3639.  Narrow-band images were
taken in December of 1991 as part of Faint Object Spectrograph General
Observer program 3194. All images were taken during the
pre-refurbishment period, when the optics were affected by spherical
aberration. Details concerning these images are given in Table 1. The
PC constitutes the high resolution component (0.044 arcsec pix$^{-1}$)
of WF/PC I and is described in detail by Griffiths \markcite{g89}
(1989).  Assuming a distance to M51 of 9.6 Mpc (Sandage and Tammann
\markcite{st74} 1974), one PC pixel corresponds to 2.0 pc. Guiding
during all observations was carried out in coarse track, with the core
of the galaxy approximately centered on chip PC6.  Pointing during the
F555W exposures was excellent, and the three frames were found to be
coincident to within 0.1 pixels.

	 All images were reduced following the procedures
outlined by Lauer \markcite{l89} (1989).  For F555W and F664N, most
cosmic rays events were removed by pixel-by-pixel comparison of
image pairs, and the few remaining obvious cosmic rays were
removed interactively.  Only single images were available in F502N and
F547M, and obvious cosmic rays were identified interactively and
removed by interpolation.

	Deconvolution of WF/PC 1 images has become the standard
procedure for extracting information on scales approaching the
diffraction limit of the uncompromised telescope optics (\eg Lauer
\markcite{l91} \etal 1991).  The F555W image was deconvolved with a
composite Point-Spread-Function (PSF) constructed using several
dedicated PSF exposures of stars taken during the period June through
September 1992. The focus changes during this period were relatively
minor (Hasan, Burrows, \& Schroeder \markcite{hbs93} 1993), and the
composite PSF has a signal-to-noise ratio considerably better than
that of the individual images. Eighty iterations of the
Lucy-Richardson deconvolution algorithm (Richardson \markcite{r72} 1972;
Lucy \markcite{l74} 1974) were applied to the coadded F555W image,
and the results are shown in Figure \ref{fig:vband} (Plate 1).

	For F547M and the narrow-band images, we used PSFs generated
using Tiny Tim v2.1 (Krist \markcite{krist92} 1992).  While the fine
structure in the observed PSF is not well modeled by any existing
software, the gross characteristics and encircled-energy profiles
shown by these models represent the PSFs quite well. Due to the
smaller number of photons in these images, significant mottling of the
deconvolved images became noticeable beyong 20 iterations, and
deconvolution was halted at this point.  Continuum levels in the
narrow-band and F547M images were measured by determining the modes of
the pixel intensity distributions in regions devoid of emission-line
flux. The suitably-scaled F547M image was then subtracted from each of
the narrow-band images. In Figure \ref{fig:narrowband} we show the
results for the central $9 \times 9$ arcseconds for F664N and F502N
after 20 Lucy-Richardson iterations for the central regions of the
galaxy.  Also shown for comparison in Figure \ref{fig:narrowband} is
the central portion of the F555W image after 80 deconvolution
iterations.  Deconvolution was carried out on much larger (512
$\times$ 512 pixels) portions of the images, and the regions shown in
Figure \ref{fig:narrowband} are unaffected by edge effects.

	The frames shown in Figure \ref{fig:narrowband} and subsequent
figures have been rotated according to the position angle of the V3
axis of the telescope at the time of observation (Table 1). The
coordinates in all cases are with respect to a point lying midway
between the two brightest features near the nucleus in the F555W image
(see below).  The spatial offsets between the F555W and F547M images
were determined by centroiding 7 stars visible in both frames,
yielding $\Delta x = 39.42 \pm 0.14$ and $\Delta y = 69.05 \pm 0.14$
pixels, where the uncertainties are errors in the mean. Due to the low
signal-to-noise ratio, offsets between the F664N, F502N, and the F547M
images (which were all taken at a single epoch) rely heavily on a
single point-source situated $\approx 30\arcsec$ from the nucleus.
However, several highly-structured regions in the field were also
cross-correlated, and the resultant offsets were found to be in good
agreement with the offsets determined using the point-source. In all
cases, the offsets were found to be $\le 1$ pixel.

\section{Discussion}

\subsection{Stars and Dust}	

	Among the more striking features in Figure \ref{fig:vband} are
the many dust lanes in and around the nucleus of M51. Some of the dust
lanes are oriented almost radially, and several filaments apparently
penetrate the stellar concentration at the center, providing evidence
of radial gas transport. The main radial flows appear to be directed
roughly along the major axis of the bar and stellar oval (Pierce
\markcite{p86} 1986).  There is some suggestion that the major northern
radial dust lane passes by the nucleus, then turns sharply to enter
the nucleus from the south, similar to the flow patterns expected if
the semimajor axis of the bar is much smaller than the Lagrangian
radius (Athanassoula \markcite{a92} 1992).  However, the overriding
impression near the nucleus is one of chaos, with dust filaments
oriented in all directions, often sinuous and intersecting with one
another. This is interesting in that, in the absence of a stirring
mechanism, a dissipative gaseous medium would be expected to settle
into a pattern of smooth, non-intersecting flow lines . Yet there
appear to be few bright (and therefore young) stars in this region to
provide the energy required to disturb such a pattern.  Indeed,
comparing with the surface density of luminous stars at the perimeter
of the oval at $\approx 7\arcsec$, star formation appears to be
somewhat suppressed near the nucleus.

	The nucleus is apparently embedded in a large, central cluster
of stars and is itself obscured by narrow, intersecting dust lanes.
That this central cluster is distinct from the disk is shown in Figure
\ref{fig:sb}, where we have plotted surface brightness profiles along
the major and minor axes of the bar. The profiles have been computed
from 3-pixel-wide cuts across the image at position angles
150$\degrees$ and 240$\degrees$, respectively.  Though we cannot
reliably decompose the cluster and disk components due to the
complexity of the dust distribution, the cluster component evidently
dominates the light profile within $\approx 1\arcsec$. The profile
least affected by obscuration seems to be the western cut along the
minor axis at P.A.  $240\deg$, which exhibits a power-law slope
$\alpha \approx -1.7$ in the region $0.5\arcsec < r < 1.5\arcsec$. For
comparison, we show a King model (dashed line) having $r_c =
0.3\arcsec$ and a concentration parameter of 100.  Constructing a
2-dimensional image of the King model shown in Figure \ref{fig:sb} and
varying the pixel location of the center, we obtain the
best visual match to the observations using a centroid located midway
along the line connecting the two brightest pixels on either side of
the main nuclear dust lane. This point is used to define the origin
for the coordinate system used in the plates.

	The King model appears to be a reasonable approximation to the
surface brightness profile of the cluster in the region least affected
by disk stars or dust.  The value of $r_c$ may be much smaller than
our best-eyeball-fit value without affecting the fit in the dust-free
region, and $r_c = 0.3\arcsec \pm 0.05\arcsec$ (= 14 pc) constitutes
an upper limit. Since the nucleus is obscured, we cannot measure the
central stellar surface brightness. Even assuming that a King model is
a valid representation for the distribution of stars in the core, the
central surface brightness predicted in Figure \ref{fig:sb} can be
regarded as no more than a lower limit. Most of the unobscured
galactic nuclei so far studied with the Planetary Camera show nuclear
cusps of varying degrees of steepness (Lauer \etal \markcite{l95}
1995), and it is entirely possible that the underlying stellar
distribution in the nucleus of M51 is also more highly concentrated
than a King model. The total luminosity within 0.3\arcsec~ predicted
by the King model is $2.3 \times 10^7$ \Lsol, and the corresponding
central luminosity density is 4200 \Lsol pc$^{-3}$.  Assuming orbital
isotropy and adopting a central velocity dispersion of 118 km s$^{-1}$
(Whitmore, McElroy, \& Tonry \markcite{wmt85} 1985), we obtain a formal
value for the $V$-band core mass-to-light ratio of 2.8 \Msol/\Lsol.
The actual value could be either higher or lower depending on the
trend of the product ($Ir$) near the origin.

  Previous reports discussing the F547M image have described the dust
lanes obscuring the nucleus as having an ``X'' configuration.  This
morphology has variously been interpreted as an edge-on disk and a
separate band of dust (Kinney \etal \markcite{k92} 1992) or possibly
two intersecting disks viewed edge-on (Cowen \markcite{c92} 1992).  This
``X''-morphology is not nearly as evident in the F555W image, and the
question arises whether this difference is due to the increase in
signal-to-noise ratio or to contamination of the F555W image by line
emission.

	In Figure \ref{fig:vbandexpand} we have subtracted the
line-emission components from the F555W image and ``unsharp-masked''
the result to enhance the contrast on small scales. The F664N and
F502N images were each scaled to the F555W exposure time, and scaled
again using the relative system throughputs at the appropriate
locations in the F555W band-pass as given in the WFPC Handbook. The
sum of these two images was then subtracted from the F555W image, and
the result was divided by a slightly smoothed version of itself. In
the region where the \ha emission is most intense (0\Sec17 southwest
of the nucleus), \ha and [OIII] emission contribute a
relatively modest 2\% and 4\% of the total flux in the F555W image,
respectively.  To within the uncertainties, this agrees with a flux
measurement made at the same position after scaling and subtracting
the F547M image from the F555W image. We conclude that the apparent
differences between the F555W and F547M images are primarily due to
differences in signal-to-noise ratio and resolution. The F547M image,
divided by a slightly smoothed version of itself, is shown in the
lower panel of Figure \ref{fig:vbandexpand} for comparison.

	In the upper panel of Figure \ref{fig:vbandexpand}, the main,
elongated, nuclear dust lane (MNDL) running east-west across the
nucleus, while roughly centered in the transverse direction, does not
appear to be centered on the nucleus along its length.  The western
end of the MNDL evidently extends at least 1\Sec1 from the center of
the central cluster, while the eastern end of the dust lane ends
fairly abruptly at $\sim$ 0\Sec4. If the MNDL were an edge-on,
equilibrium, circular disk, it would be orbiting a point at least 16
pc removed from the center of the central star cluster. This would not
be a stable configuration, and we believe that the MNDL is more likely
to be part of a tidally-sheared cloud of material that recently
entered the nuclear region on a plunging orbit, the plane of which may
be nearly coincident with our line of sight. The fact that the long
axis of the MNDL is nearly perpendicular to the radio jet (see below)
may simply be a consequence of the much shorter survival time of any
dust filaments whose orbits would carry them into the path of the jet.
Surprisingly, there appears to be no obvious connection between the
MNDL and any of the larger, extra-nuclear dust filaments.

	Closer inspection of the upper panel in Figure
\ref{fig:vbandexpand} reveals what appears to be substantial, westward
``curling'' of the secondary nuclear dust lane (the cross bar of the
``X'') both to the north and to the south of the MNDL. This would be
naturally consistent with a ring or disk of material approximately 52
pc in diameter, orbiting the nucleus, and oblique to both the radio
jet and to our line of sight.  The center of the partial ellipse
traced by the dust falls within 2 pixels of the center of the central
star cluster, and the missing northwestern portion of the ring might
simply be hidden from view by the stars in the cluster. However, the
line-subtracted F555W image (and to a lesser extent the F547 image)
shows an apparent discontinuity between the northern and southern
secondary dust lanes. There is some uncertainty in regard to the
amount of continuum light at the juncture of the MNDL and the
southern, secondary dust lane stemming from the use of different PSFs
and levels of deconvolution in the F555W and narrow-band images.
Nonetheless, no reasonable scaling of the narrow-band images can
remove the apparent discontinuity, and we conclude that the northern
and southern secondary dust lanes are probably physically distinct.
Indeed, based purely on continuity arguments, it appears in Figure
\ref{fig:vbandexpand} that the northern secondary dust lane bends
towards the south as it crosses the MNDL, giving rise to the
0\Sec2 dust feature extending south from the eastern end of the
MNDL.

\subsection{Emission-Line Regions}

	In Figure 5 we compare the spatial distributions of H$\alpha$
and 6cm radio emission with one another, and with distributions of
[OIII] and F555W flux. FCJLH first showed, based on optical and radio
luminosity and on the forbidden-line widths, that the primary site of
activity in M51 is an extra nuclear cloud (XNC) lying $\approx
3\arcsec$ south of the nucleus. Moreover, the emission-line ratios
observed in the XNC were found to be similar to those in the
shock-excited gas in some supernova remnants (SNRs), though the
minimum energy computed for the XNC is roughly two orders of magnitude
larger than that computed for the very radio-luminous Galactic SNR Cas
A. Combining this with the limb-brightened morphology of the XNC, the
inferred mass motions, and the abrupt changes in the spectral
characteristics going from the disk to the XNC, FCJLH concluded that
the gas in the XNC is excited by shocks rather than photoionization by
a central source. The scale of the phenomenon, along with the presence
of a large ring of optical emission with similar spectral
characteristics 9\arcsec~ northwest of the nucleus on a line joining
the nucleus and the XNC, suggested that the ultimate source of the
activity was in the nucleus. FCJLH proposed that a nuclear jet,
creating shocks and inflating bubbles of gas in the disk, might be the
source of the activity in the XNC.

	Cecil \markcite{c88} (1988) used Fabry-Perot observations to
investigate the kinematics of the gas in the nuclear regions of M51
and found, among other things, that the nucleus and the XNC were
connected by a tongue of high-velocity, ionized gas. He also found
that the largest range of gas velocities was to be found at the point
labeled $P$ in Figure \ref{fig:narrowband}. He largely confirmed and
extended the work of FCJLH, developing a model in which a nuclear jet,
directed into the plane of the disk (and making an angle of about
70\arcdeg~ with our line of sight), strikes the periphery of a dense
cocoon of material at an oblique angle (at $P$) and is scattered to
form a secondary shock front (the arcuate feature labeled $S$ in
Figure \ref{fig:narrowband}). Crane \& van der Hulst \markcite{cvh92}
(1992) later found a very narrow, jet-like radio feature connecting
the nucleus with $P$, strongly supporting the jet-hypothesis for the
the origin of the bubbles. The naturally-weighted, 6cm VLA radio map
of Crane \& van der Hulst is shown superimposed on the F555W and F664N
images in the upper panels of Figure \ref{fig:contour}.

	Near the nucleus, the F664N image (Figure
\ref{fig:narrowband}) reveals extended regions of relatively low-level
emission, as well as localized areas of rather more intense activity.
In Figure \ref{fig:contour} we show contours of \ha emission (as
measured from the F664N image) superimposed on F555W and F502N. The
isopleths immediately surrounding the nucleus are significantly
elongated, with the long axis pointed almost directly at $P$. The
strongest emission is found 0\Sec2 to the south of the MNDL in a
narrow, 0\Sec2-long feature aimed about 10\arcdeg~ eastwards of $P$.
To the north of the MNDL, the strongest \ha emission ($\approx 25$\%
less intense than the peak emission to the south of the MNDL) is
located just on the edge of the MNDL, within 0\Sec1 of the center of
the central star cluster. It is perhaps noteworthy that this northern
source does not lie on the line projected along the long axis of the
southern feature (the northern source is situated about 0\Sec15 too
far to the east). On the other hand, a line connecting the northern
and southern emission peaks projects to lie almost on top of $P$. The
apparent morphologies of the emission peaks may be largely a
consequence of variable extinction in the vicinity of the MNDL.
Alternatively, this may also be evidence for the presence of optical
``wobbles" in the jet of the sort seen in other active nuclei The line
connecting $P$ with the nuclear emission peaks makes an angle of
$79\arcdeg \pm 2\arcdeg$ with the long axis of the MNDL.

	The regions of peak nuclear emission in F502N correspond
almost exactly with the locations of peak emission in F664N. However,
the two F502N peaks on either side of the MNDL have very similar peak
intensities, and both peaks are relatively compact, with FWHM $\sim$
0\Sec17. The pattern of nuclear [OIII] emission appears roughly
biconical (with opening angle $\approx 45\arcdeg$), consistent with the
presence of twin ionization cones along the path of the jet. Just as
in the case of F664N, the apparent centerline of the southern
ionization cone differs by about 10 degrees from the radial passing
through $P$, though the northern and southern peaks together point
almost directly at $P$. 

	Using an aperture of radius 0\Sec1 and the system
throughput curves given in the WFPC Handbook, we find for the southern
peak F(H$\alpha$ + [NII]) = $2.5 \times 10^{-14}$ ergs s$^{-1}$
cm$^{-2}$ and F([OIII]) = $9.5 \times 10^{-15}$ ergs s$^{-1}$
cm$^{-2}$. For the northern peak, we find F(H$\alpha$ + [NII]) = $1.9
\times 10^{-14}$ ergs s$^{-1}$ cm$^{-2}$ and F([OIII]) = $9.3 \times
10^{-15}$ ergs s$^{-1}$ cm$^{-2}$. The uncertainties due to photon
statistics are $\approx 1 \times 10^{-16}$ ergs s$^{-1}$
cm$^{-2}$ in each case.  The emission at the nucleus itself is a
about a factor of two less than the emission from the peaks on either
side. If the depression is due to absorption by the MNDL and the
extinction in the $V$-band is approximately 2 magnitudes (assuming
that the King model in Figure \ref{fig:sb} gives a reasonable
indication of the underlying stellar central surface brightness), this
would be roughly consistent with the extinction properties of dust in
our own Galaxy.

	0\Sec5 north of the northern emission peak, visible in both
F664N and F502N, is the Northern Compact Nebula (NCN) first described
by Goad \& Gallagher \markcite{gg85} (1985). The region measures about
0\Sec5 $\times$ 0\Sec2 and its long axis is roughly perpendicular to
the direction of the radio jet. Two similarly-shaped, though somewhat
fainter regions are evident 1\Sec0 and 1\Sec8 north of the nucleus,
respectively. The 6cm radio contours suggest enhanced radio emission
in these regions as well. Goad \& Gallagher found that the radial
velocities measured in the NCN were about 35 km s$^{-1}$ greater than
predicted using the rotation curve of Goad, De Veny, \& Goad
\markcite{gdvg79} (1979), and that the [NII] line widths were FWZI
$\ge $300 km s$^{-1}$.  Examining both F555W and Figure
\ref{fig:vbandexpand}, we find patches or filaments of dust adjacent
to and radially outwards from each of the three regions of \ha
emission. The NCN itself may also be an extension of the northern
secondary dust lane discussed previously.  FCJLH put forth a ring of
\ha emission $9\arcsec$ north of the nucleus as evidence for a
now-extinguished, northwards-directed, nuclear jet.  The high velocity
widths in the NCN, combined with the extended radio emission to the
north, suggest that a residual jet or wind in this region continues to
interact with the ISM in the disk.

	Beyond $1\arcsec$ from the nucleus, the strongest source of
\ha and [OIII] emission is at $P$, consistent with Cecil's contention
that $P$ corresponds to the primary contact between the nuclear jet
and a relatively dense cloud of gas in the disk. Inspection of Figure
\ref{fig:contour} reveals that $P$ is indeed situated at the inner end
of a large, radially-oriented, dust filament. $P$ is resolved in both
F502N and F664N, and visibly elongated perpendicular to the direction
of the jet.  The peak [OIII] emission in the extra-nuclear cloud is
coincident (to within $\sim 1$ pixel) of the peak emission in \ha.
The integrated [OIII] flux within 0\Sec3 of the brightest pixel in $P$
is $8.6 \times 10^{-15}$ ergs s$^{-1}$ cm$^{-2}$, of which half falls
in an aperture of radius 0\Sec17 (8 pc). This is consistent with the
diameter of the radio jet (0\Sec3) estimated by Crane \& van der Hulst
\markcite{cvh92} (1992). While it is possible that the apparent size
of $P$ may be in large part due to the distribution of gas in the
disk, the similarity in cross section between $P$ and the radio
feature would be consistent with a jet which remains highly
collimated, despite its somewhat sinuous appearance in the radio map.

	The arcuate emission region southwest of $P$ (Cecil's
structure $S$) borders for some of its length a dusty arc of similar
proportions. A cut across the shock front at its narrowest point (very
near the end of the pointer indicating structure $S$ in Figure
\ref{fig:narrowband}) yields a FWHM $\approx 0\Sec3$, making it only
barely resolved. The morphology of $S$ is consistent with the
inflation model of Cecil \markcite{c88} (1988), though the features we
see probably tell us more about the distribution of gas in the disk
than about the mechanics of the shock front.

	There is a good spatial correlation (best revealed by blinking
images) between the F664N-bright regions and the radial dusty regions
visible in F555W extending along either side of the radio jet.  The
emission is distributed primarily along the inner edge of the dust,
extending from the nuclear region to well past $P$.  This implies
significant interaction between the jet and the disk material along
the whole length of the jet, and not just at $P$ and $S$. Could the
jet itself, by interacting with the gas in the disk, be responsible
for the chaotic appearance of the dust filaments in Figure 1? FCJLH
measured a total optical luminosity in the XNC of $3.6 \times 10^{41}$
ergs s$^{-1}$. Assuming that the jet has been injecting this amount of
energy into the disk over the past $10^7$ years (equivalent to one
orbital period at 10\arcsec), we obtain a total energy input of $2
\times 10^{56}$ ergs. This may be compared with the total orbital
energy within 10\arcsec~ of the nucleus of $6 \times 10^{56}$ ergs,
computed by integrating over the surface brightness profile in Figure
2 and using $v_{rot} = 14.53 R$ for $R < 9\arcsec$ (Goad, De Veny, \&
Goad \markcite{gdvg79} 1979). Since the gas accounts for only a
fraction of the total mass in this region, there is clearly more than
sufficient energy in the jet to substantially stir up the ISM.
Radiative dissipation would tend to smooth out any jet-induced
turbulence on timescales of an orbit or two, and the appearance of the
dust in the Figure 1 suggests that the stirring of the ISM by the jet
has been either ongoing or recurrent for at least the last $10^7$
years.

	Outside the nucleus, the NCN, and the XNC, we see relatively
low signal-to-noise ratio \ha enhancements, three of which are labeled
$a, b,$ and $c$ in Figure \ref{fig:narrowband}. In all cases, the
emitting regions seen in F664N correspond to regions of at least some
obscuration in F555W. Region $a$ appears in F555W to be an apparently
isolated patch of dust, while $b$ and $c$ are situated very near the
apparent intersections of two dust filaments. The former is perhaps
being illuminated by leakage of ionizing radiation from the central
source, while the latter, if the intersections of the dust filaments
are not merely chance projections, may result from the shock-induced
heating of physical collisions between filaments.

	In the upper panels of Figure \ref{fig:contour} we compare the
6cm VLA radio observations (as digitized from Crane \& van der Hulst
\markcite{cvh92} 1992) with the F664N image. The radio contours have
been shifted 0\Sec2 west and 0\Sec6 north to bring the region of peak
radio emission into agreement with the optical position of the
nucleus. The disagreement between the VLA and {\it HST} coordinates is
within local astrometric uncertainties (Evans \etal \markcite{e91}
1991). After shifting, there is good, detailed agreement between the
contours of the 6cm emission and the regions of ionization in the XNC.
The radio emission shows a relatively steep gradient in the vicinity
of the emission-line arcs, with the peak radio flux concentrated {\it
within} structure $S$. This is consistent with the identification of
$S$ with the high-temperature bow shock of an expanding shell.  On the
other hand, it appears that the optical hot spot is not a feature in
the radio emission. This would suggest that the radio emission comes
from particles from the nucleus, rather than from particles
accelerated in the post shock flow. It also suggests that these
particles are not escaping upstream of the shock, because if they
could get into the strong magnetic fields behind the radiative shocks,
we would see a strong brightening in the radio flux just outwards of
the bow shock.

	The radio contours are compared with the F555W image in
Figure \ref{fig:contour}. Given the correspondence between the radio
contours and the \ha emission, and the tendency for the
\ha emission to lie along the inner borders of some of the
dust filaments, we expect that the radio contours might also be
correlated with regions of dust obscuration. The radio contours do
indeed trace the outlines of some of the dust features south of the nucleus,
supporting the view that the bubble is interacting with the dust. 

\section{Summary}

	Broad and narrow-band Planetary Camera imaging of the nuclear
region of M51 has added support to currently-held views concerning the
source of the nuclear activity, and has raised new questions concerning
the fueling of the mini-AGN at the center. In particular, we conclude

\begin{itemize}

\item[(i)] that the distribution of small-scale dust filaments within
7\arcsec~ of the nucleus appears largely chaotic,

\item[(ii)] that the dust lane obscuring the center of M51 is extended
too far in one direction to be part of a stable torus of material
orbiting the AGN,

\item[(iii)] that the center of M51 is dominated by an apparently
discrete cluster of stars which is an order of magnitude more luminous
than the brightest Galactic globular cluster,

\item[(iv)] that the width of the [OIII] emission region in the extra-nuclear
cloud is consistent with the radio observations, which suggest a jet
which is of order 10 pc or less in diameter, and

\item[(v)] that the emission-line regions are generally confined to the
inner boundaries of dust filaments, supporting the view that the jet
is ramming through (and stirring up) the ISM in the disk.

\end{itemize}

	Multi-color and narrow-band observations with the refurbished
{\it HST} will allow a more complete study of the structure of the
nucleus of M51 and the properties of the gas and dust. Near IR
observations would be particularly useful to minimize the obscuring
effects of the dust and allow us to more closely examine the region
immediately surrounding the AGN. High-resolution, HST
spectrophotometry will be needed to establish densities, temperatures,
chemical compositions, and ionization mechanisms in the nuclear
environs.

\acknowledgments

We are grateful to Bill Mathews for useful discussions during the
analysis of these images. This research was conducted by the WF/PC
Investigation Definition Team, supported in part by NASA Grant No.
NAS5-1661.

\clearpage

\figcaption{F555W Planetary Camera image of the core region of M51
after 80 iterations of the Lucy-Richardson deconvolution algorithm.
The stretch is logarithmic, and the coordinates are seconds of arc
with respect to the assumed position of the nucleus (see text).
\label{fig:vband}}

\figcaption{Broad-band and continuum-subtracted narrow-band images of
the nuclear region of M51. The stretch in all cases is logarithmic,
and the images have been rotated so that North is up and East is to
the left.  The F555W ($V$) image has undergone 80 Lucy-Richardson
deconvolution iterations, while deconvolution of the lower
signal-to-noise ratio narrow-band images was halted at 20 iterations.
The upper left panel shows a combination of the images in the three
other panels, with the color associations as indicated. The primary
(P) and secondary (S) contacts between the nuclear flow and the disk
material (see text) are indicated in the lower-left panel, as is the
Northern Compact Nebula (NCN). Local H$\alpha$ enhancements which
appear to be related to features visible in F555W are situtated at
$a$, $b$, and $c$.  \label{fig:narrowband}}

\figcaption{$V$-band surface brightness profiles along the major and
minor axes of the central bar. The dashed line is a King model with
core radius $r_c = 0\Sec3$ and tidal radius $r_t = 30\arcsec$.
\label{fig:sb}}

\figcaption{Unsharp-masked images of the central dust lanes in M51.
For F555W (upper panel), the image shown in the upper-right panel of
Figure 2 has been emission-line subtracted using the F664N and F502N
images. For both F555W and F547M (lower panel), the images have been
divided by median-window-smoothed versions of themselves to bring out
the extended structure of the nuclear dust lanes. The stretch in both
panels extends from 0.95 to 1.05. In the upper panel, the $0\Sec5
\times 0\Sec5$ region just south of the western end of the main
nuclear dust lane is the result of an imperfectly removed flat-field
defect. Indicated are features discussed in the text, including (a)
the northern secondary dust lane, (b) the main nuclear dust lane
(MNDL), and (c) the southern secondary dust lane.
\label{fig:vbandexpand}}

\figcaption{F555W and continuum-subtracted, narrow-band images with
superimposed 6cm and \ha contours. The contours in the upper two
panels are the naturally-weighted, 6cm VLA observations of Crane \& van
der Hulst (1992).  The contours in the lower panels derive from 
a slightly smoothed version of the continuum-subtracted F664N image.
\label{fig:contour}}

\end{document}